\documentclass[journal]{IEEEtran}

\ifCLASSINFOpdf
\else
\fi

\usepackage[absolute]{textpos}
\usepackage[margin=0.553in,top=1in]{geometry}
\usepackage{amsmath, amssymb, bm, cite, epsfig}
\usepackage{epstopdf}
\usepackage{graphicx}
\usepackage{array}
\usepackage{multirow}
\renewcommand{\arraystretch}{1.5}
\usepackage{amsfonts}
\usepackage{tabularx} 
\usepackage{tabu}
\usepackage{pbox}
\usepackage{makecell}

\usepackage{booktabs}
\usepackage{ctable}
\usepackage{xcolor,colortbl}
\usepackage{supertabular,booktabs}
\usepackage{longtable}
\usepackage{tabu}
\newcommand{\squeezeup}{\vspace{-2.5mm}}
\definecolor{Gray}{gray}{0.85}
\definecolor{LightCyan}{rgb}{0.8,1,1}

\def\beq{\begin{equation}}
\def\eeq{\end{equation}}
\def\beqa{\begin{eqnarray}}
\def\eeqa{\end{eqnarray}}
\def\beqan{\begin{eqnarray*}}
\def\eeqan{\end{eqnarray*}}

\def\tm1{t\! - \! 1}
\def\tp1{t\! + \! 1}

\renewcommand\IEEEkeywordsname{Index Terms}

\begin{document}
	\begin{textblock}{18.9}(1.5,0.5)
		\centering
		\noindent\Large S. Sun, H. Yan, G. R. MacCartney Jr., and T. S. Rappaport, "Millimeter wave small-scale spatial statistics in an urban microcell scenario," \textit{2017 IEEE International Conference on Communications (ICC)}, Paris, May 2017.
	\end{textblock}
\title{Millimeter Wave Small-Scale Spatial Statistics in an Urban Microcell Scenario}

\author{\IEEEauthorblockN{Shu Sun, Hangsong Yan, George R. MacCartney Jr., and Theodore S. Rappaport}\\
\IEEEauthorblockA{NYU WIRELESS and NYU Tandon School of Engineering, New York University, Brooklyn, NY, USA 11201\\
\{ss7152,hy942,gmac,tsr\}@nyu.edu }

\thanks{Sponsorship for this work was provided by the NYU WIRELESS Industrial Affiliates program and NSF research grants 1320472, 1302336, and 1555332. The authors thank Yunchou Xing, Jeton Koka, Ruichen Wang, and Dian Yu for their help in conducting the measurements.}
}
% make the title area
\maketitle

\begin{abstract}
%This paper presents outdoor wideband small-scale spatial fading and autocorrelation measurements and results at the 73 GHz millimeter-wave (mmWave) band conducted in downtown Brooklyn, New York. Both directional and omnidirectional receiver (RX) antennas are studied. Two pairs of transmitter (TX) and RX locations were tested with one line-of-sight (LOS) and one non-line-of-sight (NLOS) location combination, where a linear track was employed at each RX to move the antenna in half-wavelength increments. Measured data reveal that the small-scale spatial fading of the received signal voltage amplitude follows log-normal distribution for the synthesized omnidirectional RX antenna pattern in the NLOS environment, otherwise the signal voltage amplitudes are generally Ricean-distributed for both omnidirectional and directional RX antenna patterns under both LOS and NLOS conditions. Sinusoidal exponential and typical exponential functions are suitable for modeling small-scale spatial autocorrelation of the received signal voltage amplitude in LOS and NLOS environments in most cases, respectively. Results herein are valuable for characterizing small-scale spatial fading and autocorrelation properties in multiple-input multiple-output (MIMO) systems for fifth-generation (5G) cellular communications, especially for mmWave frequencies.
This paper presents outdoor wideband small-scale spatial fading and autocorrelation measurements and results in the 73 GHz millimeter-wave (mmWave) band conducted in downtown Brooklyn, New York. Both directional and omnidirectional receiver (RX) antennas are studied. Two pairs of transmitter (TX) and RX locations were tested with one line-of-sight (LOS) and one non-line-of-sight (NLOS) environment, where a linear track was employed at each RX to move the antenna in half-wavelength increments. Measured data reveal that the small-scale spatial fading of the received signal voltage amplitude are generally Ricean-distributed for both omnidirectional and directional RX antenna patterns under both LOS and NLOS conditions in most cases, except for the log-normal distribution for the omnidirectional RX antenna pattern in the NLOS environment. Sinusoidal exponential and typical exponential functions are found to model small-scale spatial autocorrelation of the received signal voltage amplitude in LOS and NLOS environments in most cases, respectively. Furthermore, different decorrelation distances were observed for different RX track orientations, i.e., for different directions of motion relative to the TX. Results herein are valuable for characterizing small-scale spatial fading and autocorrelation properties in multiple-input multiple-output (MIMO) systems for fifth-generation (5G) mmWave frequencies.
\end{abstract}


\IEEEpeerreviewmaketitle
\section{Introduction}
The millimeter-wave (mmWave) bands hold promising prospects for providing substantially broader bandwidths and higher data rates required for the fifth-generation (5G) wireless communications~\cite{Rap15}. There have been intensive investigations on mmWave propagation characteristics and channel modeling in various environments~\cite{Rap_2013_millimeter,Rap_2015_Wideband,Ghosh_2014_Millimeter,Mac_2015_Indoor,Samimi_2016_3D,Han16_Outdoor}, which is paving the way for 5G mmWave system design and deployment. Nevertheless, the knowledge on some aspects of mmWave propagation is still scarce, such as small-scale spatial fading and autocorrelation properties of received signal voltage amplitudes, which are crucial in multiple-input multiple-output (MIMO) channel modeling for 5G wireless communications.

Previously, indoor continuous-wave (CW) propagation experiments and analysis at 910 MHz were performed in~\cite{Bul}, which showed that both temporal and spatial signal envelope fading distributions were Riciean with $K$-factors of 6 dB to 12 dB and 2 dB, respectively~\cite{Bul}. Indoor corridor measurements were conducted to study small-scale fading characteristics at 15 GHz~\cite{Wang_2015}, which demonstrated that small-scale fading in indoor corridor scenarios could be well described by Ricean distributions with $K$-factors ranging from 5 dB to 10 dB for most measurement locations when omnidirectional antennas were used at both the transmitter and receiver, while the value of Ricean $K$-factors increased significantly when employing directional antennas. Another measurement campaign carried out in an industrial environment showed the applicability of the Rayleigh distribution for modeling the small-scale fading for all time excess delay bins except the first one that was inclined to be Nakagami-$m$ fading~\cite{Tanghe_2009_Statistical}, whereas~\cite{Rappaport_91_TCOM} found that log-normal distribution was a better fit over all time bins. The impact of bandwidth on the small-scale fade depth was explored in~\cite{Malik_2007_Impact}, indicating a decrease in the fade depth as the bandwidth increased, attaining 4 dB fade depth at 1 GHz. For a bandwidth exceeding 1 GHz, the fade depth became less dynamic over bandwidth. The work in~\cite{Holtzman94} and~\cite{Durgin00} also showed decreased fading depths as the signal bandwidth became larger. Durgin \textit{et al.}~\cite{Durgin98} noticed that at 1.92 GHz fade depths would decrease when more directional antennas were used. Henderson \textit{et al.} examined a 2.4 GHz indoor channel and tried to find an appropriate small-scale fading distribution~\cite{Henderson_2008_finding}. Three types of distributions were utilized to fit the measured data: Rayleigh, Ricean, and Two-Wave-Diffuse-Power (TWDP) distributions, among which the Ricean distribution had the most extensive applicability for describing small-scale fading characteristics in most indoor cases~\cite{Henderson_2008_finding}. Furthermore, unified probability density function (PDF) formulas for characterizing a wide range of small-scale fading distributions were derived in~\cite{S2010corrections, chai2009unified}. 

%The small-scale fading distributions included in the unified formula were multiple-waves-plus-diffuse-power (MWDP) fading, $\kappa - \mu$, Nakagami-$m$, Ricean (Nakagami-$n$), Nakagami-\textit{q} (Hoyt), Rayleigh, Weibull, and $\alpha-\mu$ distributions~\cite{S2010corrections, chai2009unified}.

In this paper, we investigate the small-scale spatial fading and autocorrelation behavior of the received signal voltage amplitude for both directional and omnidirectional antenna patterns in an urban microcell (UMi) environment for both line-of-sight (LOS) and non-line-of-sight (NLOS) conditions, based on a set of outdoor measurements conducted in downtown Brooklyn, New York, using a stationary 35.31-cm (about 87 wavelengths at 73.5 GHz) linear track at the receiver to move the directional antenna, with a carrier frequency of 73.5 GHz and a radio frequency (RF) bandwidth of 1 GHz.

\section{73 GHz Small-Scale Fading Measurements}
\subsection{Measurement Hardware}
The small-scale measurement campaign was conducted with a wideband sliding correlator channel sounder that transmitted at a center frequency of 73.5 GHz~\cite{Cox72a,Mac14a}. The sounding sequence at the transmitter (TX) was a pseudorandom noise (PN) sequence of length 2047 and was generated with a field-programmable-gate-array (FPGA) and high-speed digital-to-analog converter (DAC) at a rate of 500 Megachips-per-second (Mcps)~\cite{Mac16a}. The baseband sequence was mixed with a 5.625 GHz intermediate-frequency (IF) and then upconverted with a 67.875 GHz local oscillator (LO) to reach a center RF of 73.5 GHz~\cite{Mac16c}. The 1 GHz RF null-to-null bandwidth signal centered at 73.5 GHz was then transmitted through a steerable high-gain pyramidal horn antenna with 27 dBi gain and 7$^\circ$ azimuth and elevation half-power beamwidth (HPBW) that was elevated to a height of 4 m above ground.

At the receiver (RX), the received signal was captured with a steerable widebeam horn antenna at a height of 1.4 m with 9.1 dBi gain and 60$^\circ$ HPBW in the azimuth and elevation planes. The wideband RF signal was then downconverted with a 67.875 GHz LO and then further demodulated by a 5.625 GHz IF into its in-phase ($I$) and quadrature-phase ($Q$) baseband voltage signals, typical of a super-heterodyne architecture. The $I$ and $Q$ voltage signals were then amplified and correlated in analog with a signal identical to the TX PN sequence, but at a slightly offset rate of 499.9375 Mcps. This sliding correlation method resulted in a slide factor of $8\,000$, also known as the time dilation factor~\cite{Rap02aChap5}. After the analog correlation, the $I$ and $Q$ channel correlated voltages were sampled with a high-speed oscilloscope and then squared and summed together ($I^2+Q^2$) in software to result in the recorded power delay profile (PDP). For recording PDPs at various angles, the TX and RX antennas were rotated by FLIR gimbals with LabVIEW software~\cite{Mac14a}. Similarly, the RX antenna was translated in linear directions along a track controlled by LabVIEW software~\cite{Mac16a,Deng16a}. Other specifications about the measurement hardware are detailed in Table~\ref{tbl:hardware}. 

\subsection{Measurement Environment and Procedure}
\begin{figure}
	\centering
	\includegraphics[width=3.2in]{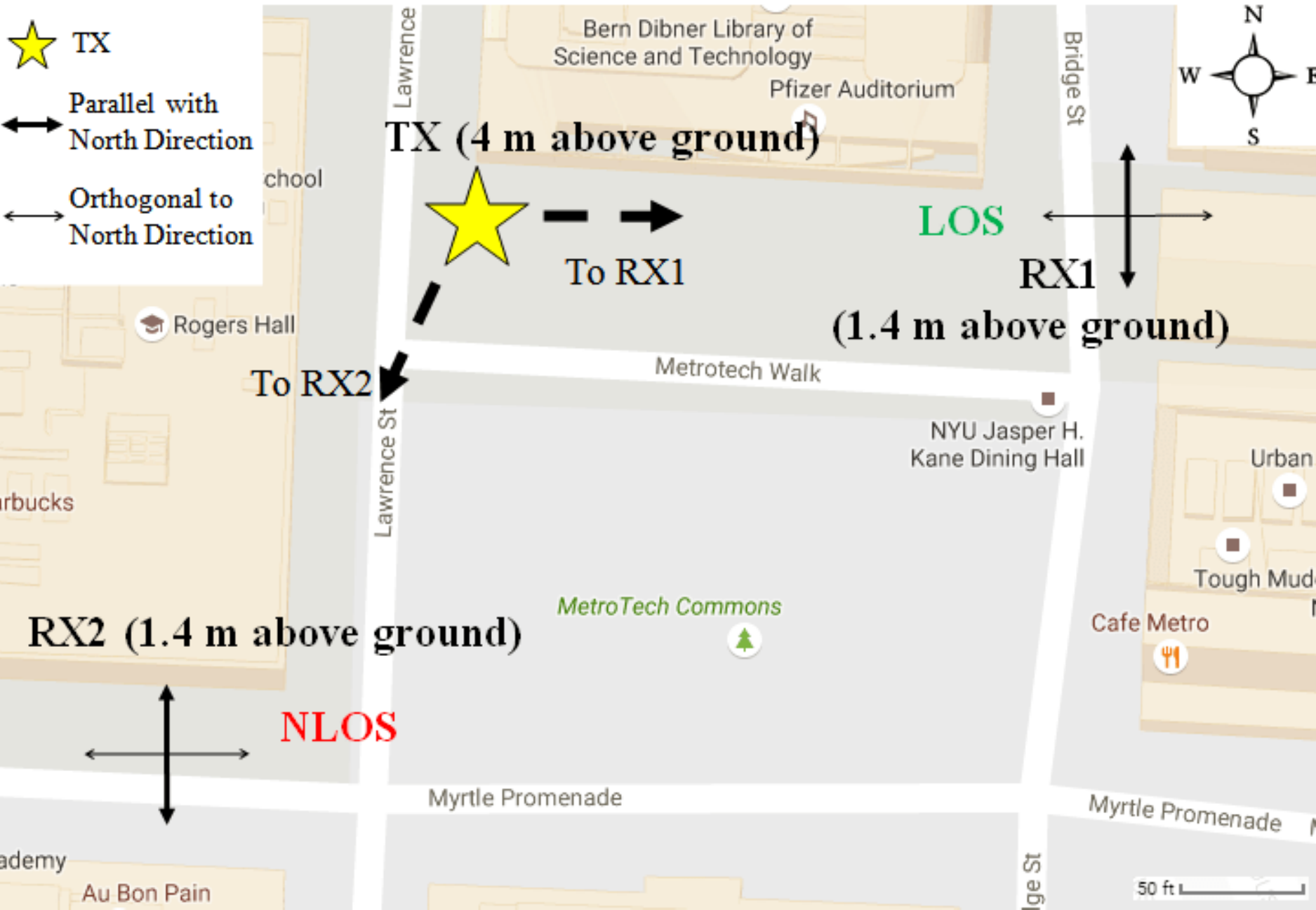}
	\caption{2D map depicting the small-scale measurement environment and the TX and RX locations.}
	\label{fig:TX_RX_Location}
\end{figure}

The small-scale linear track measurements at 73 GHz were performed on the campus of NYU Tandon School of Engineering, representative of a UMi environment. The measurement environment is detailed in Fig.~\ref{fig:TX_RX_Location} with illustrations of the TX and RX locations. One TX location with the antenna height set to 4.0 m above the ground and two RX locations with the antenna height set to 1.4 m were selected to perform the measurements, where one RX was LOS to the TX while the other was NLOS. The TX was placed near the southwest corner of the Dibner library building (north and center in Fig.~\ref{fig:TX_RX_Location}), the LOS RX was located 79.9 m away from the TX, and the NLOS RX was shadowed by the southeast corner of a building (Rogers Hall on the map) with a T-R separation distance of 75.0 m. 

\begin{figure}[b]
	\centering
	\includegraphics[width=2.5in]{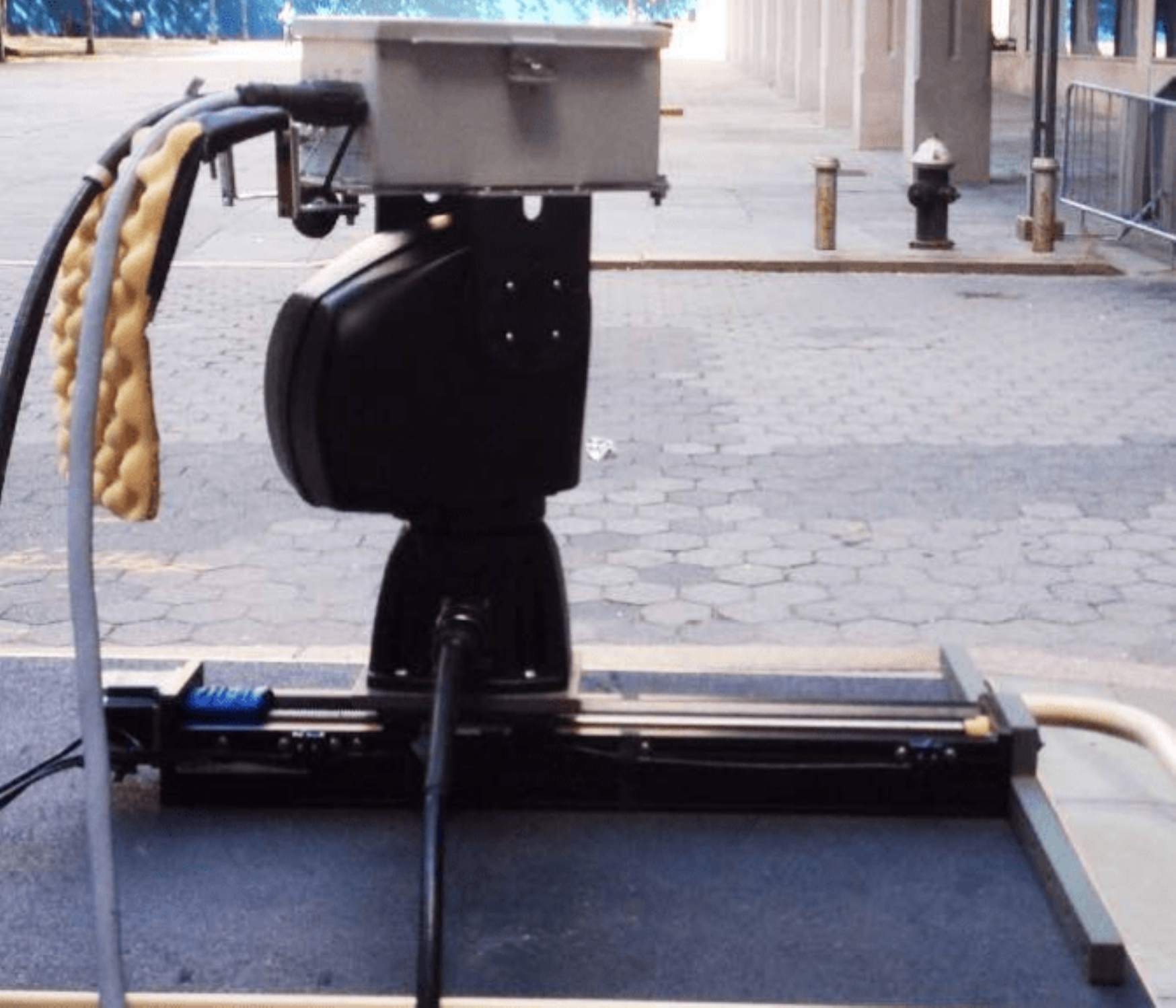}
	\caption{The 35.31-cm (about 87 wavelengths) linear track used in the 73 GHz small-scale spatial fading and correlation measurements.}
	\label{fig:Track}
\end{figure}

A stationary 35.31-cm (about 87 wavelengths at 73.5 GHz) linear track was used at each RX location in the measurements (as shown in Fig.~\ref{fig:Track}), over which the RX antenna was linearly-moved in increments of half-wavelength (2.04 mm) for 175 track positions. Two orientations of the linear track were tested in the measurements: orthogonal and parallel to the initial RX antenna azimuth pointing angle. For each track orientation, six sets of small-scale fading measurements were performed, where the elevation angle of the RX antenna remained fixed at 0$^\circ$ (parallel to horizon) with a different azimuth angle fixed for each set of the measurements where the adjacent azimuth angles were separated by 60$^\circ$ (HPBW increments), such that the RX antenna swept over the entire azimuth plane after rotating through the six pointing angles. The TX antenna elevation angle was always fixed at 0$^\circ$ (parallel to horizon). Under the LOS condition, the TX antenna was pointed at 90$^\circ$ in the azimuth plane, directly towards the RX location; for NLOS, the TX antenna azimuth pointing angle was 200$^\circ$, roughly towards the southeast corner of Rogers Hall in Fig.~\ref{fig:TX_RX_Location}.

As a comparison, the 28 GHz small scale measurements presented in~\cite{Samimi_2016_28} investigated the small scale fading and autocorrelation of \textit{individual multipath} voltage amplitudes using a 30$^\circ$ Az./El. HPBW RX antenna, whereas this paper studies 73 GHz fading and autocorrelation using a wider HPBW (60$^\circ$) RX antenna, and focuses on received signal voltage amplitude by integrating the area under the \textit{entire PDP} curve and then taking the square root of the total power, instead of individual multipath voltage amplitude. 

%the square root of the total received power obtained as the integral of the \textit{entire PDP}, 

\begin{table}
	\renewcommand{\arraystretch}{1.3}
		\caption{Hardware Specifications of Small-Scale Fading and Correlation Measurements}~\label{tbl:hardware}
		\fontsize{8}{8}\selectfont
		\begin{center}
		%\begin{tabular}{|>{\centering\arraybackslash}m{3.0cm}|>{\centering\arraybackslash}m{2.9cm}|>{\centering\arraybackslash}m{3.5cm}|>{\centering\arraybackslash}m{2.5cm}|>{\centering\arraybackslash}m{2.5cm}|>{\centering\arraybackslash}m{0.6cm}|}\hline
		\begin{tabular}{|>{\centering\arraybackslash}m{3.5cm}|>{\centering\arraybackslash}m{3.5cm}|}\hline
		
			\textbf{Description} & \textbf{Specification}  \\ \hline \hline
			\textbf{Broadcast Sequence}			        & 11\textsuperscript{th} order PN Code (L = $2^{11}-1$ = 2047)   \\ \hline
			\textbf{TX and RX Antenna Type}             & Rotatable Pyramidal Horn Antenna  \\ \hline
			\textbf{TX Chip Rate} 			            & 500 Mcps		            \\ \hline
			\textbf{RX Chip Rate}                       & 499.9375 Mcps            \\ \hline
			\textbf{Slide Factor $\gamma$}              & $8\,000$                    \\ \hline
			\textbf{RF Null-to-Null Bandwidth} 	    & 1 GHz			           \\ \hline
			%\textbf{PDP Detection}			            & \multicolumn{2}{c|}{FFT matched filter}		   \\ \hline
			%\textbf{sampling rate} 			            & \multicolumn{2}{c|}{1.5 GS/s on I and Q} 			 \\ \hline
			%\textbf{Multipath Time Resolution}          & \multicolumn{2}{c|}{2 ns}     \\ \hline
			%\textbf{Minimum Periodic PDP Interval}      & \multicolumn{2}{c|}{32.752$\mu$s}                 \\ \hline
			%\textbf{Maximum periodic PDP records per snapshot}       & \multicolumn{2}{c|}{41000 PDPs}      \\ \hline
			\textbf{PDP Threshold}                      & 20 dB down from max peak  \\ \hline
			\textbf{TX/RX Intermediate Frequency}       & 5.625 GHz          \\ \hline
			\textbf{TX/RX Local Oscillator}             & 67.875 GHz (22.625 GHz$\times 3$)   \\ \hline
			\textbf{Carrier Frequency}                  & 73.5 GHz        \\ \hline
			\textbf{TX Power}                           & 14.2 dBm   \\ \hline
			\textbf{TX Antenna Gain}                    & 27 dBi  \\ \hline
			\textbf{TX Azimuth/Elevation HPBW}      &  $7^\circ$/$7^\circ$  \\ \hline
			%\textbf{TX Antenna Polarization}            & V          \\ \hline
			\textbf{EIRP}                               & 41.2 dBm  \\ \hline
			\textbf{TX Heights}                         & 4.0 m   \\ \hline
			\textbf{RX Antenna Gain}                    & 9.1 dBi  \\ \hline
			\textbf{RX Azimuth/Elevation HPBW}      & $60^\circ$/$60^\circ$  \\ \hline
			\textbf{TX-RX Antenna Polarization}            & V-V (Vertical-to-Vertical)  \\ \hline
			\textbf{RX Heights}                         & 1.4 m     \\ \hline
			\textbf{Maximum Measurable Path Loss}       & 168 dB    \\ \hline
		\end{tabular}
	\end{center}
\end{table}


\subsection{Data Post-Processing}
The processing gain incurred by sliding correlation and amplifiers in the measurement system was extracted in post-processing for each individual PDP. A double-threshold --- 20 dB down from the maximum peak and 5 dB up from the mean noise floor --- was applied to each PDP to ensure signal integrity by excluding random noise spikes and low signal-to-noise ratio (SNR) artifacts. In this work, small-scale statistics for the directional measurements are meant for the amplitude of the received signal by taking the square root of the total received power in the linear scale obtained as the integral of the \textit{entire PDP}, instead of for individual \textit{multipath components}. The total received power for each PDP from directional measurements was determined by summing up the total power (above the threshold) under the PDP curve in the linear scale. The approximated omnidirectional received power was synthesized from directional measurements using the approach presented in~\cite{Sun_2015_Synthesizing}, and was named omnidirectional received power. Although the RX antenna did not sweep the entire $4\pi$ Steradian sphere but just the azimuth plane spanning $\pm 30^{\circ}$ elevation coverage with respect to the horizon, a majority of the arriving energy was captured as verified in~\cite{Sun_2015_Synthesizing}.

\section{Omnidirectional Small-Scale Spatial Statistics}
As described above, a rotatable directive horn antenna was used at the RX side to obtain directional PDPs in the small-scale spatial fading and autocorrelation measurements. In channel
modeling, however, omnidirectional statistics are usually preferred, since arbitrary antenna patterns can be implemented according to one\rq{}s own needs based on the omnidirectional channel model. Therefore, in this section, we investigate small-scale spatial fading and autocorrelation of the received signal voltage amplitude for the omnidirectional RX antenna pattern~\cite{Sun_2015_Synthesizing}, based on the measured data at the 73 GHz band.

\subsection{Omnidirectional Small-Scale Spatial Fading}
\begin{figure}
	\centering
	\includegraphics[width=3.1in]{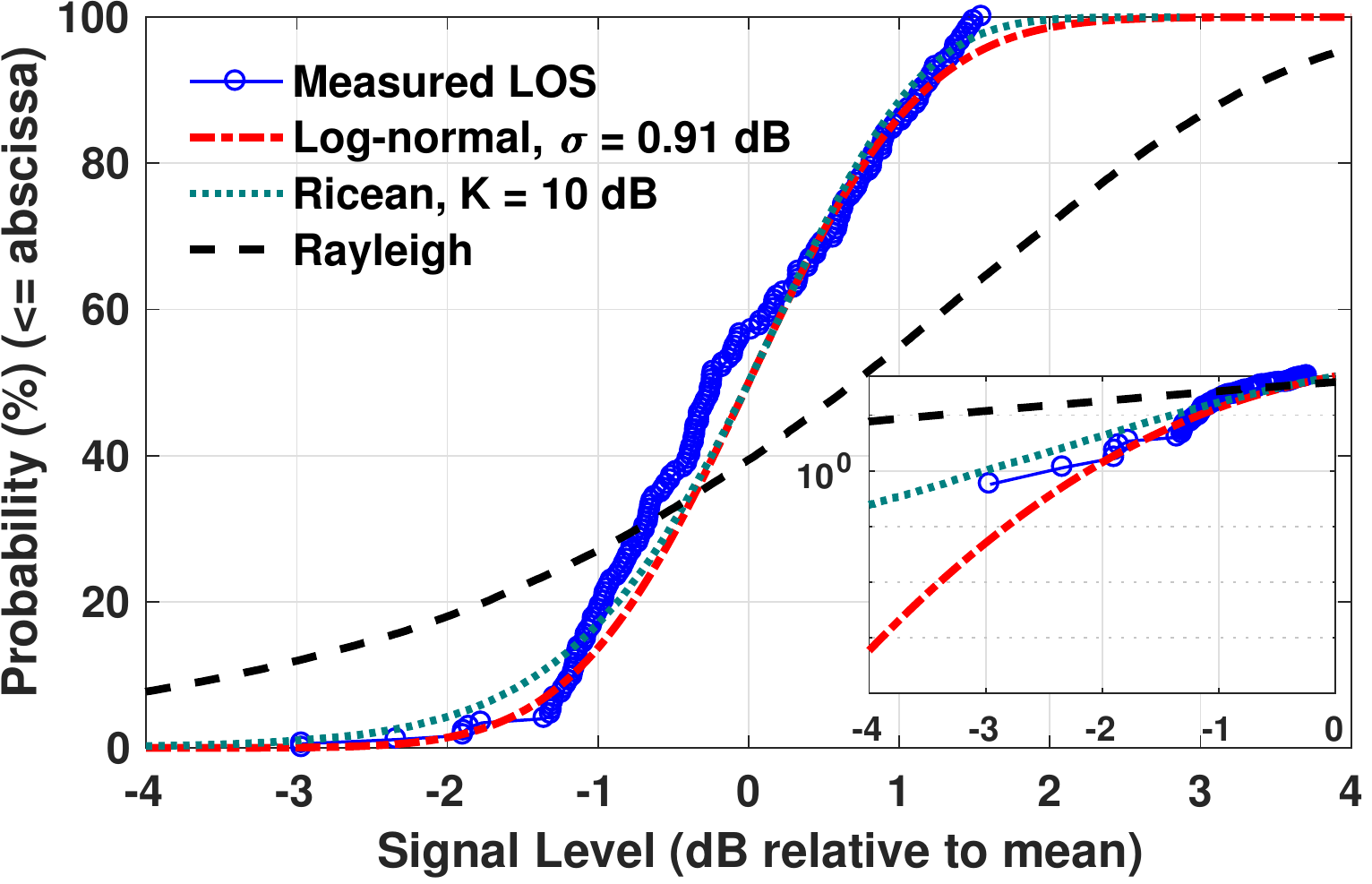}
	\caption{CDF of the measured 73 GHz broadband small-scale spatial fading distribution of the received signal voltage amplitude over a 35.31-cm (about 87 wavelengths) linear track for the omnidirectional RX antenna pattern in the LOS environment.}
	\label{fig:omniLOSFading}
\end{figure}

Fig.~\ref{fig:omniLOSFading} illustrates the cumulative distribution function (CDF) of the measured small-scale spatial fading of the received signal voltage amplitude, relative to the mean value for the omnidirectional RX antenna pattern in the LOS environment, where the track orientation is orthogonal to the direct line connecting the TX and RX. Superimposed with the measured curve are the CDFs of the Rayleigh distribution, the zero-mean log-normal distribution with a standard deviation of 0.91 dB (obtained from the measured data), and the Ricean distribution with a $K$-factor of 10 dB obtained from the measured data by dividing the total received power contained in the LOS path by the power contributed from all the other reflected or scattered paths. As shown by Fig.~\ref{fig:omniLOSFading}, the measured 73 GHz small-scale spatial fading in the LOS environment can be approximated by the Ricean distribution with a $K$-factor of 10 dB, indicating that there is a dominant path (i.e., the LOS path) contributing to the total received power, and that the received signal voltage amplitude varies little over the 35.31-cm (about 87 wavelengths) length track. The log-normal distribution does not fit the measured data well in the regions of -3 to -2.5 dB and 1.2 to 1.5 dB about the mean. The maximum fluctuation of the received signal voltage amplitude is merely 3 dB relative to the mean value, whereas Rayleigh distributed fades are much deeper. The physical reason for the tiny fluctuation of the received signal voltage amplitude about the mean level in the LOS environment can be attributed to the presence of a dominant LOS path. 

%The small-scale spatial fading distribution for the track for a parallel translation relative to the T-R line can also be described by the Ricean distribution, albeit with a much larger $K$-factor of approximately 22 dB and even tinier fluctuations of -1 dB to +0.5 dB, implying that the received power is more steady for parallel translations with respect to the direct T-R line, as compared to the orthogonal small-scale translation with respect to the T-R line. 

\begin{figure}
	\centering
	\includegraphics[width=3.1in]{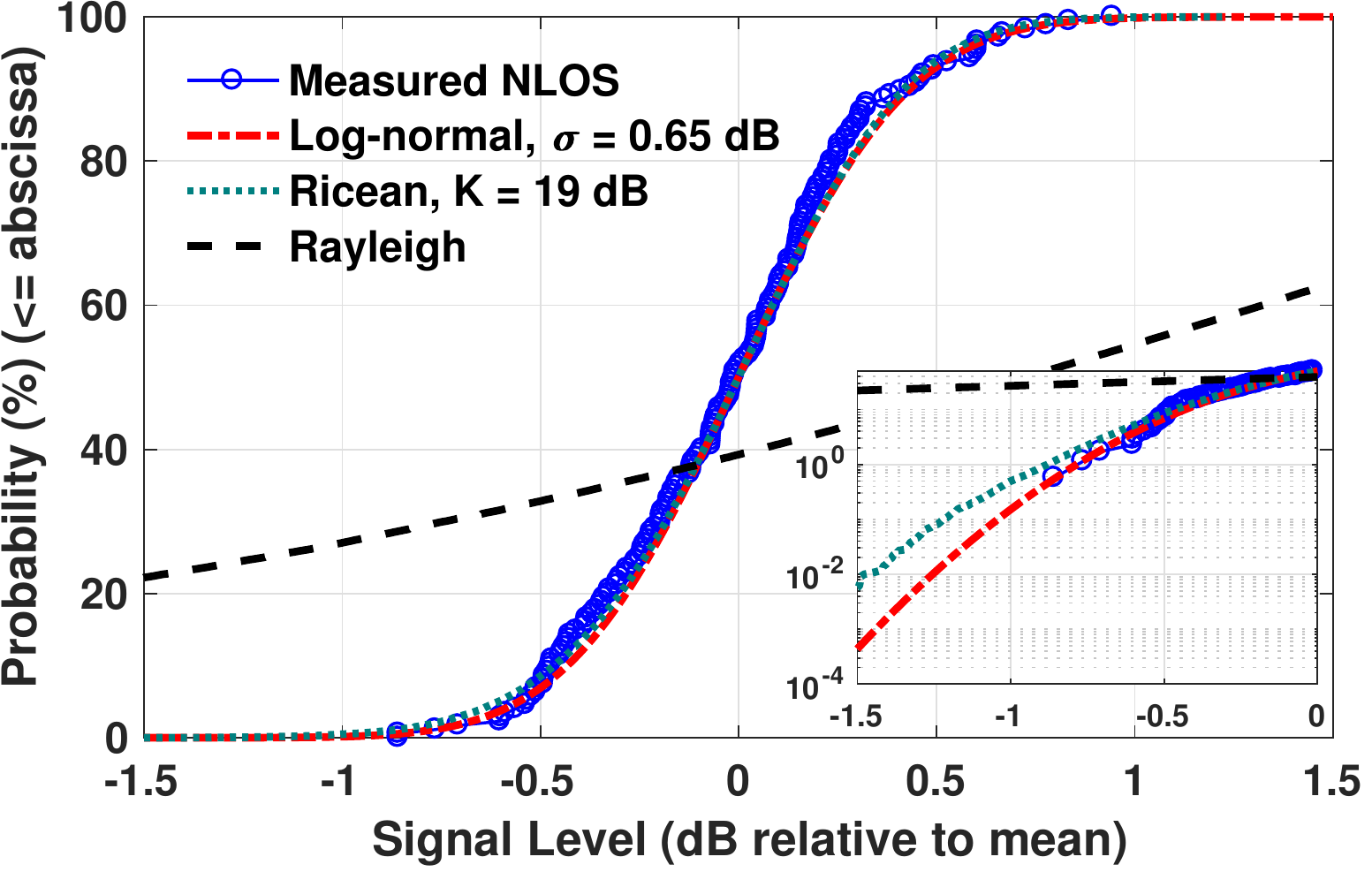}
	\caption{CDF of the measured 73 GHz broadband small-scale spatial fading distribution of the received signal voltage amplitude over a 35.31-cm (about 87 wavelengths) linear track for the omnidirectional RX antenna pattern in the NLOS environment.}
	\label{fig:omniNLOSFading}
\end{figure}

Small-scale spatial fading in the NLOS environment for the omnidirectional RX antenna pattern with the RX antenna translating from south to north is depicted in Fig.~\ref{fig:omniNLOSFading}, and the zero-mean log-normal distribution with a standard deviation of 0.65 dB (obtained from the measured data) is selected to fit the measured result, and Ricean and Rayleigh distributions are also given as a reference. As evident from Fig.~\ref{fig:omniNLOSFading}, the measured NLOS small-scale spatial fading distribution matches the log-normal fitted curve almost perfectly. In contrast, the Ricean distribution with $K$ = 19 dB does not fit the measured data as well as the log-normal distribution in the tail region around -0.6 dB to -0.8 dB of relative signal level (as shown by the inset in Fig.~\ref{fig:omniNLOSFading}), since the Ricean $K$ = 19 dB distribution predicts more occurrences of deeper fading events, whereas the log-normal distribution with a 0.65 dB standard deviation predicts a more compressed fading range of -0.8 dB to 0.8 dB about the mean, which was observed for the wideband NLOS signals. The fact that the local fading of received signal voltage amplitudes in the NLOS environment is log-normal instead of Rayleigh or Ricean is similar to models in~\cite{Rappaport_91_TCOM} for indoor factory channels. It is noteworthy that the standard deviation is only 0.65 dB, which is small and indicates minimal fluctuation in the received signal level. For the NLOS environment, there may not be a dominant path, yet the transmitted broadband signal experiences frequency-selective fading (which happens when the signal bandwidth is larger than the coherence bandwidth of the channel), different frequency components of the signal hence experience uncorrelated fading, thus it is highly unlikely that all parts of the signal will simultaneously experience a deep fade. As a consequence, the total received power (and thus voltage amplitude) does not vary significantly over a small-scale local area. This is a distinguishing feature of the wideband signal compared to its narrowband counterpart, and also echoes the work in~\cite{Holtzman94} and~\cite{Durgin00} which showed how the fading depths became smaller as the signal bandwidth increased. 

%\begin{figure}
	%\centering
	%\includegraphics[width=3.5in]{omniNLOSFlu.pdf}
	%\caption{Comparison of signal voltage amplitude fluctuation between Ricean fading, log-normal fading, and the measured 73 GHz broadband small-scale spatial fading for the omnidirectional RX antenna pattern in the NLOS environment.}
	%\label{fig:omniNLOSFlu}
%\end{figure}

\subsection{Omnidirectional Small-Scale Spatial Autocorrelation}
Apart from small-scale spatial fading, small-scale spatial autocorrelation is also a relevant research topic. Spatial autocorrelation is a metric to characterize the autocorrelation of the voltage amplitudes of the received signals across uniformly separated track positions. Spatial autocorrelation coefficient functions can be calculated using Eq.~\eqref{sac}, where $X_k$ denotes the $k\textsuperscript{th}$ linear track position, $E[~]$ is the expectation operator where the expectation is taken over all the positions $X_k$, and $\Delta X$ represents the spacing between different antenna positions on the track. 
\begin{figure*}
	\begin{equation}\label{sac}
	\rho=\frac{E\big[\big(A_k(X_k)-\overline{A_k(X_k)}\big)\big(A_k(X_k+\Delta X)-\overline{A_k(X_k+\Delta X)}\big)\big]}{\sqrt{E\big[\big(A_k(X_k)-\overline{A_k(X_k)}\big)^2\big]E\big[\big(A_k(X_k+\Delta X)-\overline{A_k(X_k+\Delta X)}\big)^2\big]}}
	\end{equation}
\end{figure*}

\begin{figure}
	\centering
	\includegraphics[width=3.0in]{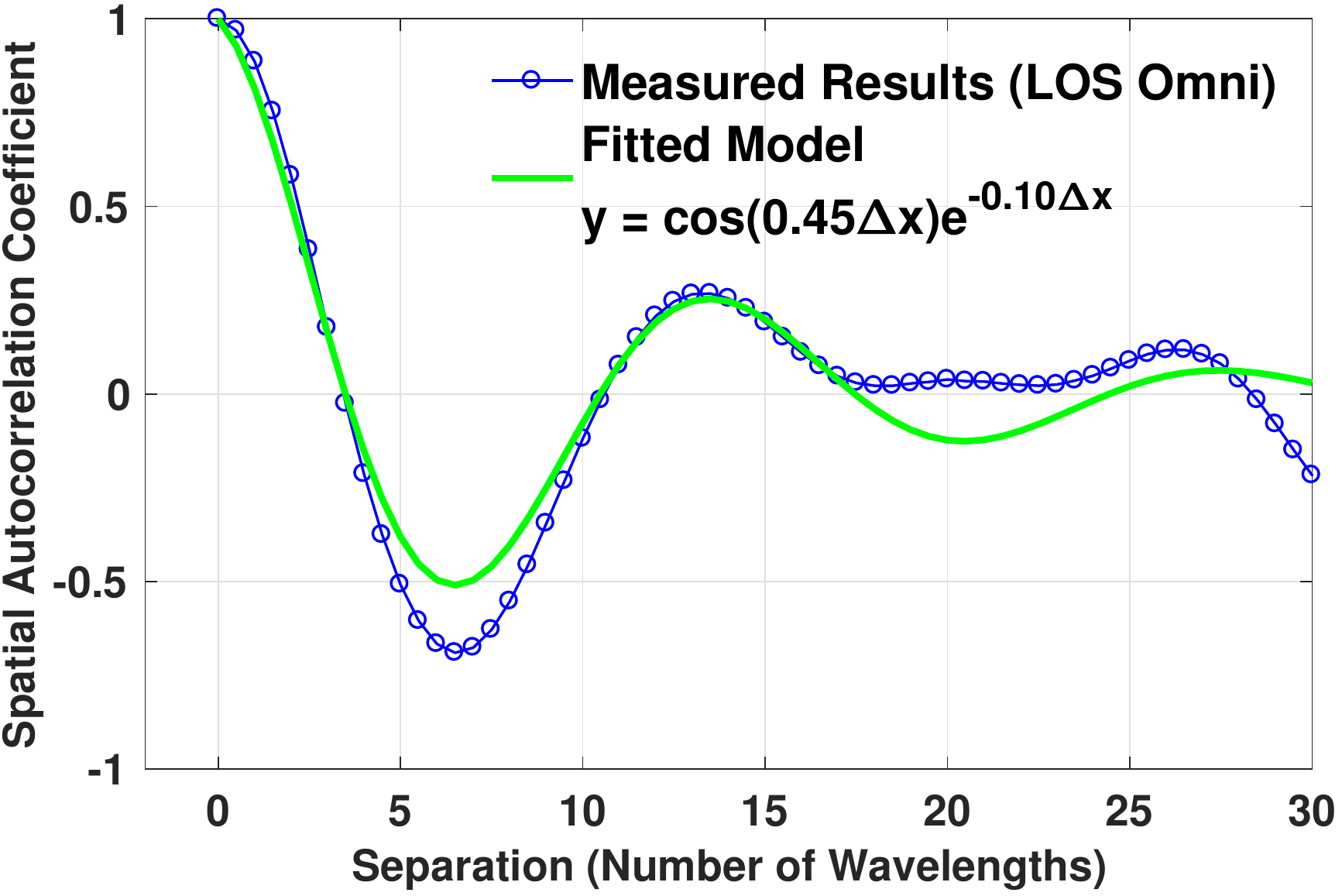}
	\caption{Measured 73 GHz wideband spatial autocorrelation coefficients of the received signal voltage amplitude over a 35.31-cm (about 87 wavelengths) linear track for the omnidirectional RX antenna pattern in the LOS environment, and the corresponding fitted model.}
	\label{fig:omniLOSCor}
\end{figure}

\begin{figure}
	\centering
	\includegraphics[width=3.0in]{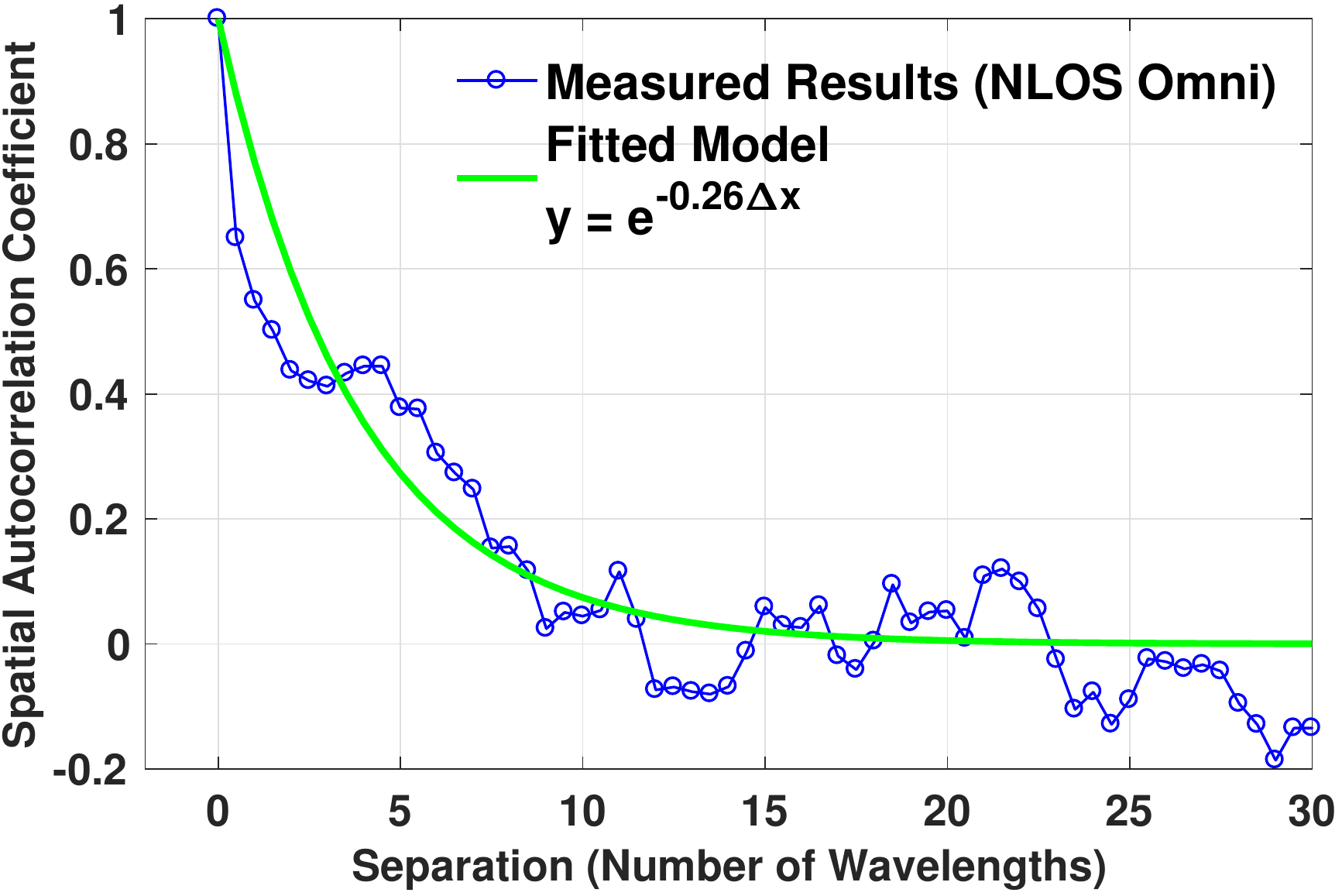}
	\caption{Measured 73 GHz wideband spatial autocorrelation coefficients of the received signal voltage amplitude over a 35.31-cm (about 87 wavelengths) linear track for the omnidirectional RX antenna pattern in the NLOS environment, and the corresponding fitted model.}
	\label{fig:omniNLOSCor}
\end{figure}

The measured 73 GHz wideband spatial autocorrelation coefficients of the received signal voltage amplitudes in LOS and NLOS conditions with the RX antenna moving from south to north are depicted in Fig.~\ref{fig:omniLOSCor} and Fig.~\ref{fig:omniNLOSCor}, respectively. Note that a total of 175 linear track positions were measured during the measurements, yielding a maximum spatial separation of 174 half-wavelengths. Only up to 30 wavelengths, however, are shown herein because little change is found thereafter and it provides at least 100 autocorrelation data points for all spatial separations on a single track, thus improving reliability of the statistics. Per Fig.~\ref{fig:omniLOSCor}, the received omnidirectional signal voltage amplitude first becomes uncorrelated at a spatial separation of about 3.5$\lambda$ (where $\lambda$ represents the wavelength), then becomes slightly anticorrelated for separations of 3.5$\lambda$ to 10$\lambda$, and becomes slightly correlated for separations between 10$\lambda$ and 18$\lambda$, and decays towards 0 sinuisoidally after 18$\lambda$. Therefore, the spatial autocorrelation can be modeled by a ``damped oscillation'' function of \eqref{expFit}~\cite{Zhang08}
\begin{equation}\label{expFit}
f(\Delta X) = \cos(a\Delta X)e^{-b\Delta X}
\end{equation}

\noindent where $\Delta X$ denotes the space between antenna positions, $a$ is an oscillation distance with units of radians/$\lambda$ (wavelength), $T=2\pi/a$ can be defined as the spatial oscillation period with units of $\lambda$ or cm, and $b$ is a spatial decay constant with units of $\lambda^{-1}$ whose inverse $d=1/b$ is the decorrelation distance with units of $\lambda$. $a$ and $b$ are obtained using the minimum mean square error (MMSE) method to find the best fit between the empirical spatial autocorrelation curve and theoretical model given by~\eqref{expFit}. The ``damped oscillation'' pattern can be explained by superposition of multipath components with different phases at different linear track positions. As the separation distance of linear track positions increases, the phase differences among individual multipath components will oscillate as the separation distance of track positions increases due to alternating constructive and destructive combining of the multipath phases. This ``damped oscillation'' pattern is obvious in LOS environment where phase difference among individual multipath component is not affected by shadowing effects that occurred in NLOS environments. The form of~\eqref{expFit} also guarantees that the spatial autocorrelation coefficient is always 1 for $\Delta X = 0$, and converges to 0 when $\Delta X$ approximates infinity. 

The spatial autocorrelation curve for NLOS environment in Fig.~\ref{fig:omniNLOSCor} exhibits a different trend from that in Fig.~\ref{fig:omniLOSCor}, which is more akin to an exponential distribution without damping, but can still be fitted using Eq.~\eqref{expFit} with $a$ set to 0. The constants $a$, $b$, are provided in Table~\ref{tbl:ModelPara}, where $T$ is the spatial oscillation period, and $d$ represents the spatial decay constant. From Fig.~\ref{fig:omniNLOSCor} and Table~\ref{tbl:ModelPara} it is clear that after 1.57 cm (3.85 wavelengths at 73.5 GHz) in the NLOS environment, the received voltage amplitudes become uncorrelated (the correlation coefficient decreases to 1/e~\cite{Samimi_2016_3D}). We note that Samimi~\cite{Samimi_2016_28} found \textit{individual multipath} voltage amplitudes in the directional measurements using a 30$^\circ$ Az./El. HPBW antenna became uncorrelated at physical distances of 0.52 cm (0.48 wavelengths at 28 GHz) and 0.67 cm (0.62 wavelengths at 28 GHz) in LOS and NLOS environments, respectively, whose decorrelation distances were smaller as compared to the \textit{total received voltage} (square root of area under the PDP) at 73 GHz using a 60$^\circ$ Az./El. HPBW antenna for both the directional antenna pattern and synthesized omnidirectional antenna pattern. 

\begin{table}[t!]
	\renewcommand{\arraystretch}{1.3}
	\caption{Spatial correlation model parameters in~\eqref{expFit} for 73 GHz, 1 GHz RF bandwidth ($\lambda$=0.41 cm) based on the fitted curves on measured data over a 35.31-cm (about 87 wavelengths) linear track.}~\label{tbl:ModelPara}
	\fontsize{8}{6.5}\selectfont
	\scriptsize
	\begin{center}\scalebox{0.90}{
			\squeezeup
			\begin{tabular}{|>{\centering\arraybackslash}m{1.6cm}|>{\centering\arraybackslash}m{1.1cm}|>{\centering\arraybackslash}m{1.8cm}|>{\centering\arraybackslash}m{0.9cm}|>{\centering\arraybackslash}m{1.8cm}|>{\centering\arraybackslash}m{0.6cm}|}\hline
				\textbf{Condition} & $\boldsymbol{a}~(rad/\lambda)$ & $\boldsymbol{T=2\pi/a}$	        & \textbf{b}~($\lambda^{-1}$)	 & $\boldsymbol{d=1/b}$             \\ \hline \hline
				\textbf{LOS Omnidirectional} & {0.45} & {14.0$\lambda$ (5.71 cm)} & {0.10} & {10.0$\lambda$ (4.08 cm)}\\ \hline
				\textbf{NLOS Omnidirectional} & {0} & {-} & {0.26} & {3.85$\lambda$ (1.57 cm)}\\ \hline
				\textbf{LOS Directional} & {0 - 0.50} & {12.6$\lambda$ - $\infty$ (5.14 cm - $\infty$)} & {0.005 - 0.195} & {5.13$\lambda$ - 200$\lambda$ (2.09 cm - 81.6 cm)}\\ \hline
				\textbf{NLOS Directional} & {0} & {-} & {0.04 - 1.49} & {0.67$\lambda$ - 25.0$\lambda$ (0.27 cm - 10.2 cm)}\\ \hline
		\end{tabular}}
	\end{center}
\end{table}

\section{Directional Small-Scale Spatial Statistics}
 This section is dedicated to the small-scale spatial fading and autocorrelation of the received signal voltage amplitudes associated with the directional RX antenna pattern. 
 
\subsection{Directional Small-Scale Spatial Fading}
\begin{figure}
	\centering
	\includegraphics[width=3.2in]{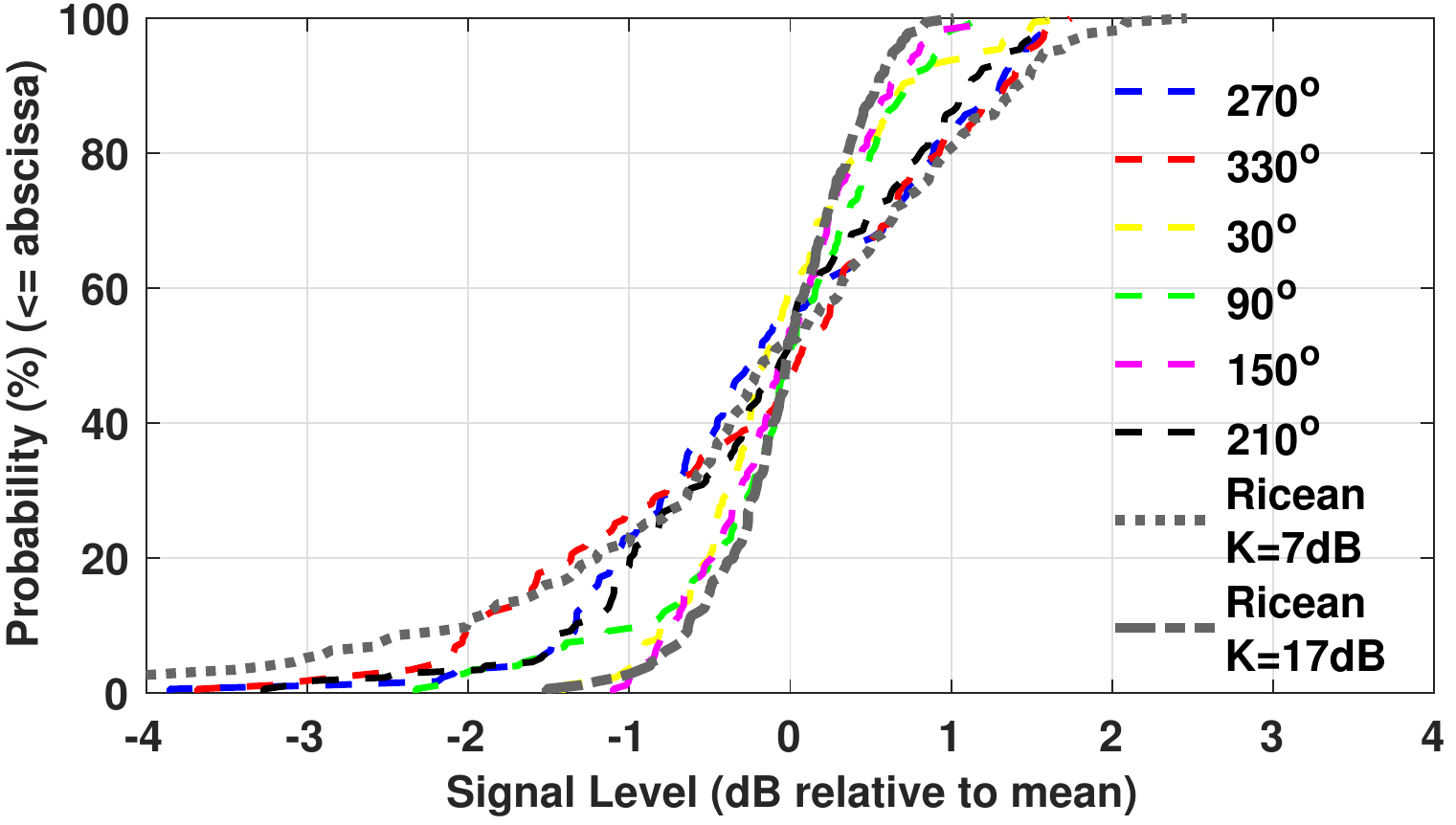}
	\caption{Measured 73 GHz small-scale spatial fading distributions of the directional received signal voltage amplitude over a 35.31-cm (about 87 wavelengths) linear track in LOS conditions, and the corresponding Ricean fitted curves with different $K$-factors. The angles in the legend denote the receiver antenna azimuth pointing angle during the small-scale measurements with $270^\circ$ denoting the angle pointing directly at the transmitter.}
	\label{fig:dirLOSFading}
\end{figure}

\begin{figure}
	\centering
	\includegraphics[width=3.2in]{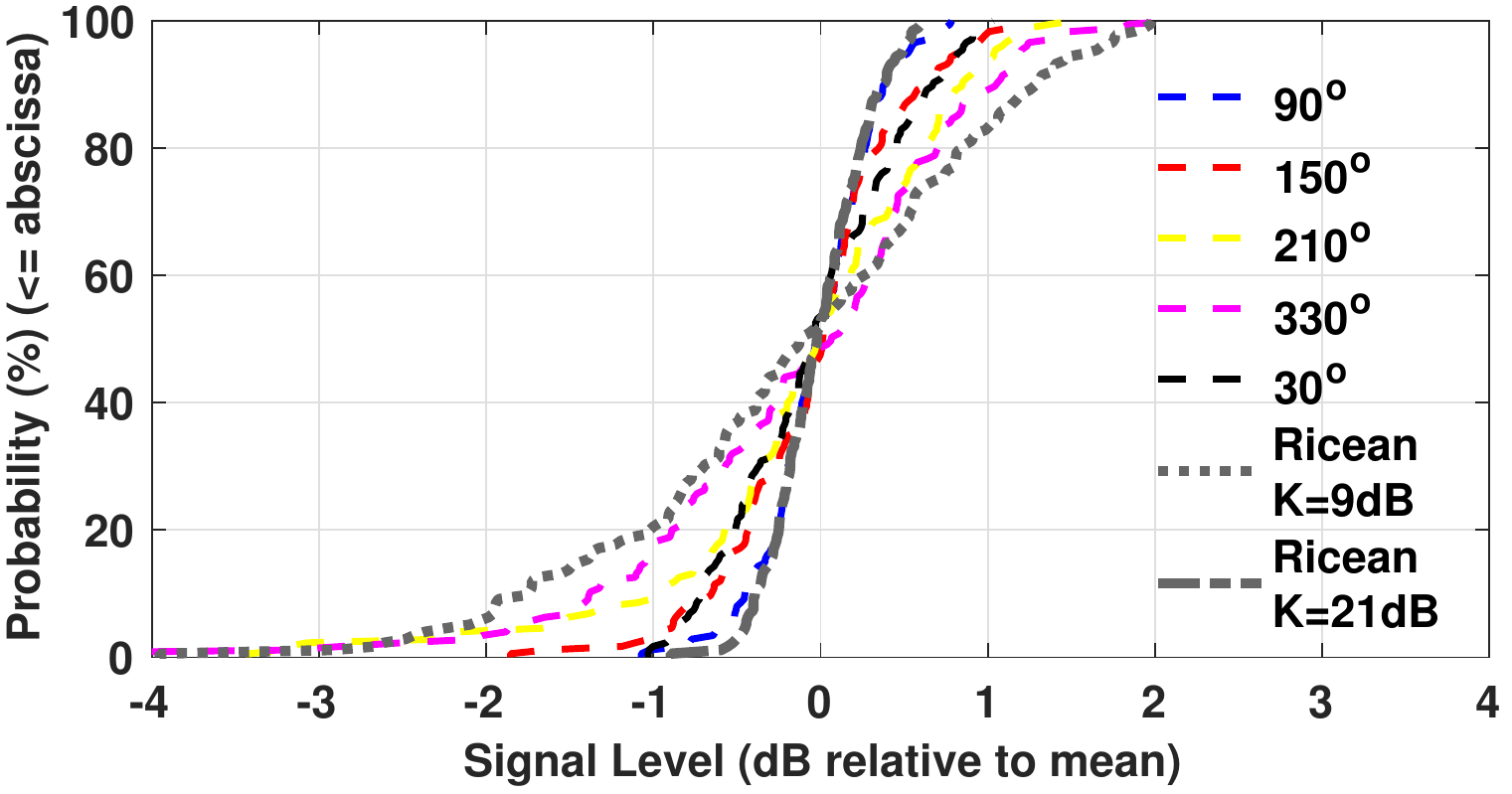}
	\caption{Measured 73 GHz small-scale spatial fading distributions of the directional received signal voltage amplitude over a 35.31-cm (about 87 wavelengths) linear track in NLOS conditions, and the corresponding Ricean fitted curves with different $K$-factors. The angles in the legend denote the receiver antenna azimuth pointing angle during the small-scale measurements with $30^\circ$ denoting the angle pointing roughly directly at the transmitter.}
	\label{fig:dirNLOSFading}
\end{figure}

The small-scale spatial fading of the received signal voltage amplitudes along the linear track using the $7^{\circ}$ azimuth and elevation HPBW TX antenna and $60^{\circ}$ HPBW RX antenna with the RX antenna translating from south to north in LOS and NLOS conditions are drawn in Fig.~\ref{fig:dirLOSFading} and Fig.~\ref{fig:dirNLOSFading}, respectively, where each measured curve corresponds to a unique RX antenna azimuth pointing angle as specified in the legend. Note that there was no signal for the RX azimuth pointing angle of 270$^{\circ}$ in the NLOS environment, thus the corresponding results are absent in Fig.~\ref{fig:dirNLOSFading}. As is observed from Figs.~\ref{fig:dirLOSFading} and~\ref{fig:dirNLOSFading}, the CDFs of measured directional spatial signal voltage amplitude fading resemble Ricean distributions in both LOS and NLOS environments, which are similar to the results obtained from 28 GHz small-scale fading measurements and results presented in~\cite{Samimi_2016_28}. The possible physical reason for such distributions is that only one dominant path (accompanied with several weaker paths in some cases) was obtained by the horn antenna due to its directionality (i.e., limited HPBW), and that mmWave propagation is directional and the channel is sparse. The Ricean $K$-factor ranges from 7 dB to 17 dB for LOS conditions, and 9 dB to 21 dB for NLOS cases, as shown in Figs.~\ref{fig:dirLOSFading} and~\ref{fig:dirNLOSFading}. The comparable Ricean $K$-factors in LOS and NLOS environments indicate that the environment (LOS or NLOS) does not influence the Ricean fading, namely, for a directional RX antenna in either LOS or NLOS environment, there exists a dominant path whose power is much higher than the other weaker paths, and the power ratio of the dominant path to weaker paths is similar in LOS and NLOS environments in mmWave channels. 

\subsection{Directional Small-Scale Spatial Autocorrelation}
%Figs.~\ref{fig:dirLOSCor} and~\ref{fig:dirNLOSCor} illustrate the spatial autocorrelation distributions of the received signal amplitudes for individual antenna pointing angles in LOS and NLOS environments, respectively. The angles in the legend denote the RX horn antenna azimuth pointing angle relative to true north. As shown by Fig.~\ref{fig:dirLOSCor}, all of the six spatial autocorrelation curves in the LOS environment exhibit sinusoidally exponential decaying trends, albeit with different oscillation and decay rates. Compared with the omnidirectional case displayed in Fig.~\ref{fig:omniLOSCor}, it is obvious that for the LOS environment, the spatial autocorrelation function of both omnidirectional and directional received signal amplitude obeys similar distribution, namely, the  sinusoidal-exponential function. On the other hand, the spatial autocorrelation curves for the directional received signal amplitude shown in Fig.~\ref{fig:dirNLOSCor} are also in line with that given by Fig.~\ref{fig:omniNLOSCor}, which follow the exponential function and agree well with the 28 GHz NLOS small-scale spatial autocorrelation distribution of multipath component amplitudes described in~\cite{Samimi_2016_28}.

\begin{figure}
	\centering
	\includegraphics[width=3.2in]{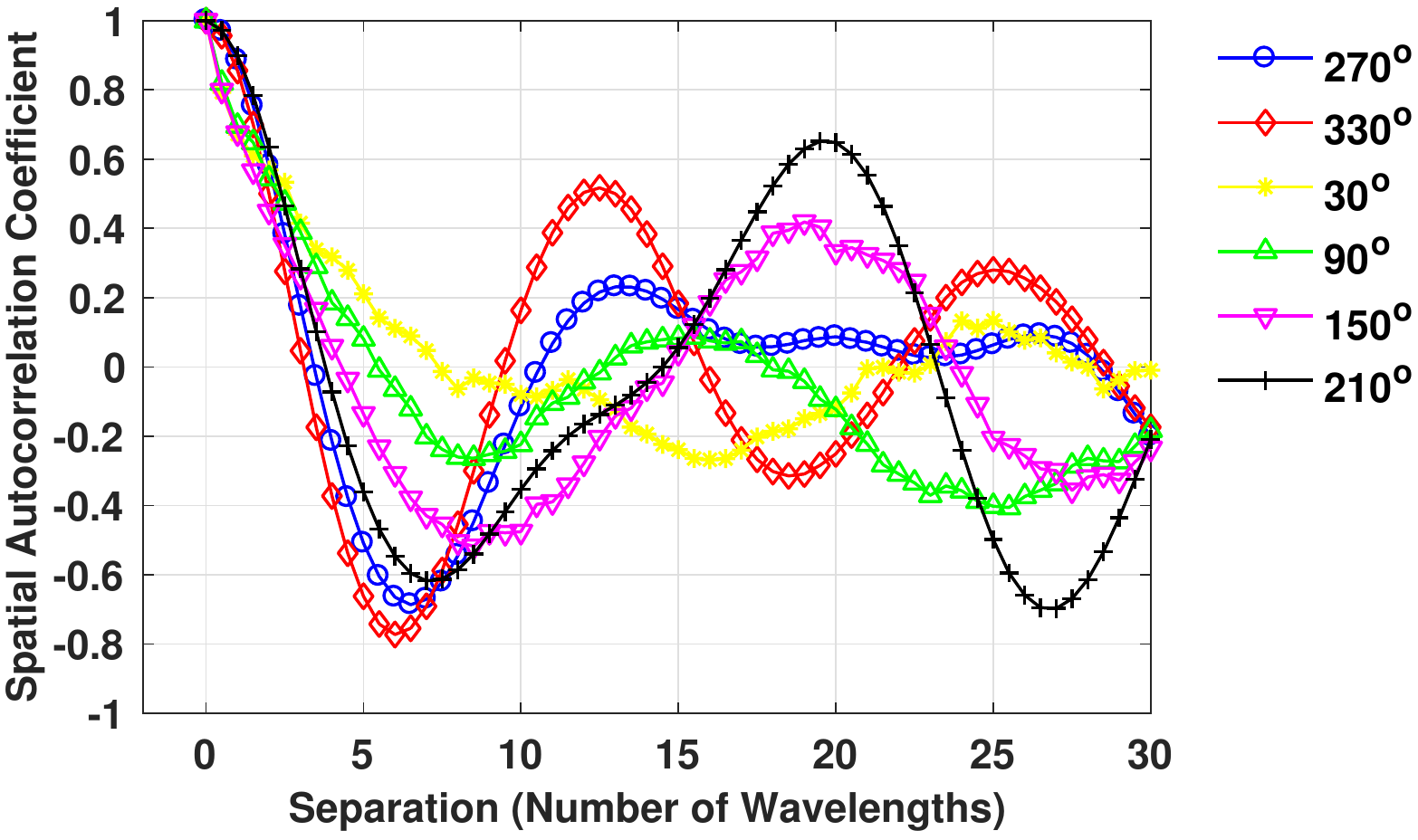}
	\caption{Measured 73 GHz spatial autocorrelation coefficients of the directional received signal voltage amplitude over a 35.31-cm (about 87 wavelengths) linear track in the LOS environment with the RX antenna translating orthogonally relative to the T-R line. The angles in the legend denote the receiver antenna azimuth pointing angle during the small-scale measurements with $270^\circ$ denoting the angle pointing directly at the transmitter.}
	\label{fig:dirLOSCor}
\end{figure}

Figs.~\ref{fig:dirLOSCor} illustrates the spatial autocorrelation coefficients of the received signal voltage amplitudes versus RX antenna separation distance for individual antenna pointing angles in the LOS environment, where the RX antenna was moved in an orthogonal manner relative to the T-R line. The angles in the legend denote the RX antenna azimuth pointing angles relative to true north. As shown by Fig.~\ref{fig:dirLOSCor}, all of the six spatial autocorrelation curves in the LOS environment exhibit sinusoidally exponential decaying trends (albeit with different oscillation and decay rates), which is similar to the omnidirectional case displayed in Fig.~\ref{fig:omniLOSCor}.

\begin{figure}
	\centering
	\includegraphics[width=3.2in]{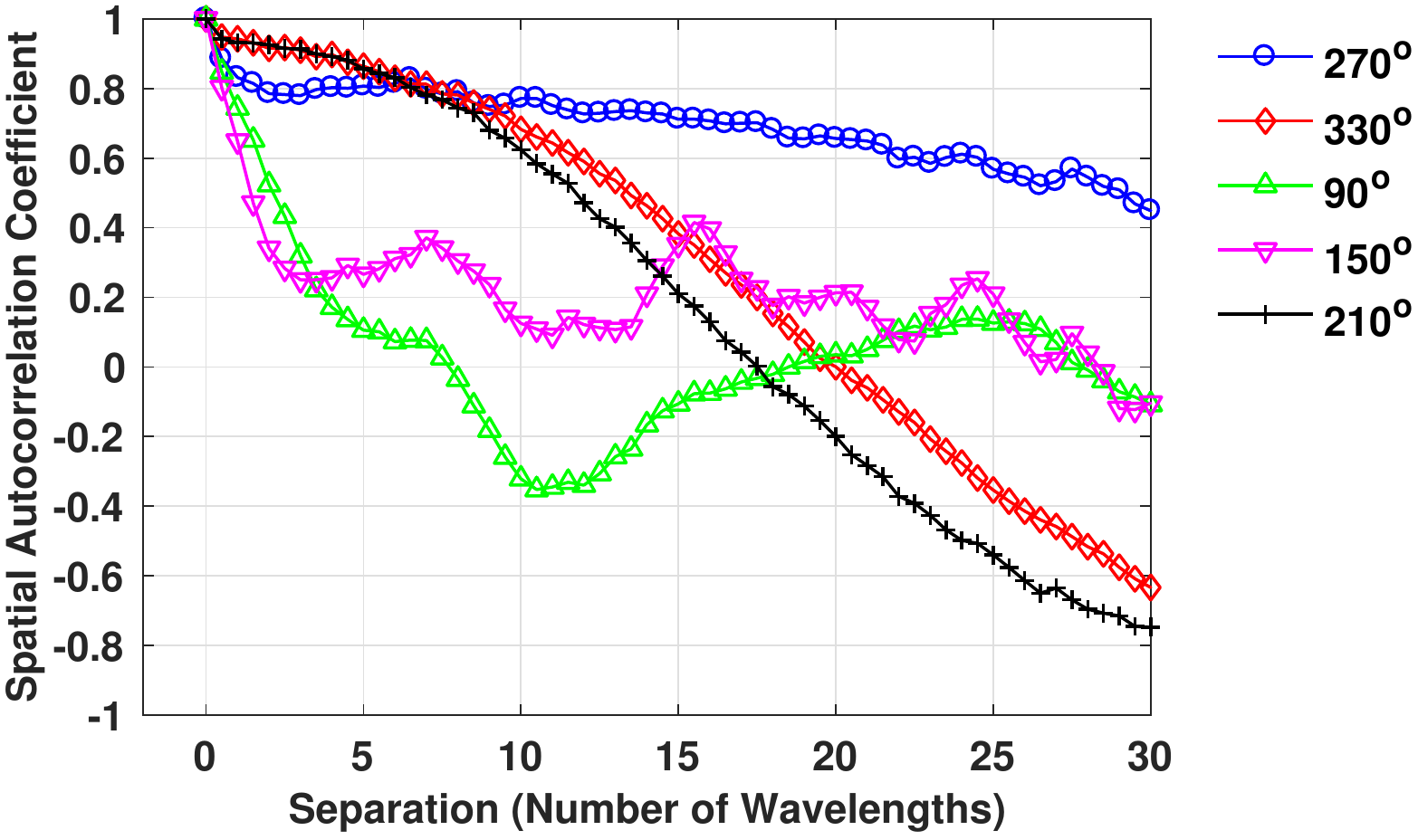}
	\caption{Measured 73 GHz spatial autocorrelation coefficients of the directional received signal voltage amplitude over a 35.31-cm (about 87 wavelengths) linear track in the LOS environment with the RX antenna translating in a parallel manner with respect to the T-R line. The angles in the legend denote the receiver antenna azimuth pointing angle during the small-scale measurements with $270^\circ$ denoting the angle pointing directly at the transmitter.}
	\label{fig:dirLOSCor_Track2}
\end{figure}

The spatial autocorrelation curves for the RX antenna translations parallel with respect to the T-R line are illustrated in Fig.~\ref{fig:dirLOSCor_Track2} (since there were random pedestrians blocking the RX antenna when the RX antenna pointed at 30$^\circ$, the corresponding result was biased hence not plotted). It is intriguing to observe that the autocorrelations do not always follow the sinusoidal-exponential function as in Fig.~\ref{fig:dirLOSCor}. For the RX antenna pointing angle of 270$^\circ$, for instance, the received signal voltage amplitude is highly correlated with a slowly-decaying autocorrelation curve, which can be physically interpreted by the fact that the TX and RX antennas were always directly pointing at each other while the RX antenna moved along the track, thus only a LOS path was received whose amplitude varied gradually with the minor changes in the T-R separation distance. In contrast, the autocorrelation function drops quickly from 1 to 0 and becomes increasingly negative as the antenna separation increases for the RX antenna pointing angles of 330$^\circ$ and 210$^\circ$, for which the RX antenna was pointing towards a building and a grove area, respectively, and there was only one resolvable multipath component in the PDP. The spatial decorrelation distance $d=1/b$ using model~\eqref{expFit} is found to be 200 wavelengths, which is an "extrapolation" of the sinusoidal model, meaning that the correlation drops to 1/e at 200 wavelengths and is always below 1/e afterwards. 

%It is possible that the LOS path and a reflected path were received simultaneously and the two paths gradually became out-of-phase from in-phase as the RX antenna moved along the track, such that the resultant total received power decreased with the displacement of the RX antenna. 

\begin{figure}
	\centering
	\includegraphics[width=3.2in]{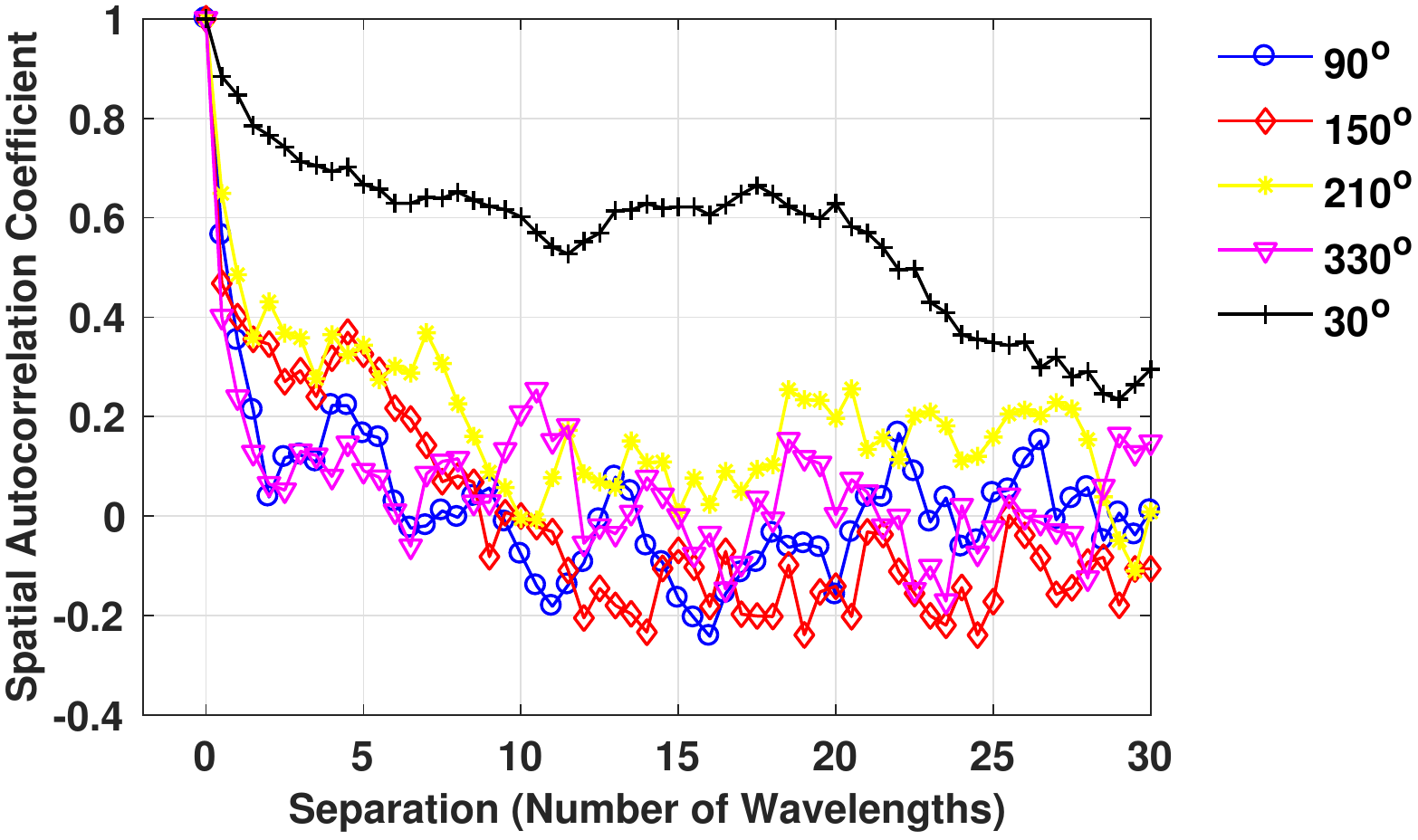}
	\caption{Measured 73 GHz spatial autocorrelation coefficients of the directional received signal voltage amplitude over a 35.31-cm (about 87 wavelengths) linear track in the NLOS environment. The angles in the legend denote the receiver antenna azimuth pointing angle during the small-scale measurements with $30^\circ$ denoting the angle pointing roughly directly at the transmitter.}
	\label{fig:dirNLOSCor}
\end{figure}

For the NLOS environment, the spatial autocorrelation coefficients of the received signal voltage amplitudes for individual antenna pointing angles for one antenna translation direction are plotted in Fig.~\ref{fig:dirNLOSCor} (the results for the other track orientation are very similar to the ones shown here), wherein the results for the majority of the RX pointing angles are in line with that given by Fig.~\ref{fig:omniNLOSCor}, which follow the exponential function and agree well with the 28 GHz NLOS small-scale spatial autocorrelation coefficients of multipath voltage amplitudes described in~\cite{Samimi_2016_28}. One exception is the 30$^\circ$ pointing angle, where the autocorrelation coefficient slowly decreases from 1 to 0.2 or so, probably due to the presence of a dominant path with a relatively constant signal level, which may be caused by the diffraction of the southeast corner of Rogers Hall in Fig.~\ref{fig:TX_RX_Location}. 

The spatial autocorrelation model parameters in~\eqref{expFit}, along with the spatial oscillation period $T$ and spatial decay constant $d$, are summarized in Table~\ref{tbl:ModelPara} for both omnidirectional and directional RX antenna patterns under LOS and NLOS environments. For the LOS environment, the decorrelation distance ranges from around five wavelengths to 200 wavelengths (2.09 - 81.6 cm), while the NLOS decorrelation distance is between 0.67 and 25 wavelengths at 73.5 GHz (0.27 - 10.2 cm) at 73.5 GHz. The 200-wavelength decorrelation distance is calculated/predicted using the model in~\eqref{expFit} based upon the measured data over a track length of 30 wavelengths, which corresponds to the $210^\circ$ and $330^\circ$ curves in Fig~\ref{fig:dirLOSCor_Track2}, in which case the RX antenna was pointing at a main reflector and there was only one resolvable multipath component in the PDP (with a 20 dB down threshold~\cite{Samimi_2016_3D}). The 200-wavelength decorrelation distance means that the correlation is smaller than 1/e after 200 wavelengths, but the correlation may also be less than 1/e at some points, and increases and decreases in a sinusoidal manner within 200 wavelengths. The above observations indicates that the decorrelation distance is the largest when the RX antenna points at a major reflector and moves in a parallel manner with respect to the T-R line, and the smallest decorrelation distance occurs when the RX antenna is pointing roughly to the opposite direction of the TX and without major reflectors. These decorrelation distances are valuable to antenna design~\cite{Ertel}.

\section{Conclusion}
%This paper has investigated the wideband small-scale spatial fading and autocorrelation of the received signal amplitudes at 73 GHz, for both omnidirectional and directional receiver antenna patterns in both LOS and NLOS environments in a UMi scenario. Small-scale spatial fading of 1 GHz bandwidth signals follows the Ricean distribution at both 28 GHz and 73 GHz for the voltage amplitude received by directional antennas, yet log-normally distributed path amplitudes were observed for the omnidirectional antenna pattern in the NLOS environment. Sinusoidal exponential and exponential functions accurately model the small-scale spatial autocorrelation of 1 GHz bandwidth signals at 73 GHz. Table~\ref{tbl:ModelPara} shows the oscillation distance/period and spatial decay constant to represent the autocorrelation, where rapid decorrelation of received voltage amplitudes occurred over 0.67 to 33.3 wavelengths (0.27 cm to 13.6 cm). The short correlation distance in general is favorable for spatial multiplexing in MIMO, since it allows for uncorrelated spatial data streams to be transmitted from closely-spaced (a fraction to several wavelengths) antennas.

This paper has investigated the wideband small-scale spatial fading and autocorrelation of the received signal voltage amplitudes at 73 GHz, for both omnidirectional and directional receiver antenna patterns in both LOS and NLOS environments in a UMi scenario. 

When measured over a 35.31-cm (about 87 wavelengths) linear track, for the omnidirectional antenna pattern, the LOS fading obey the Ricean distribution with a $K$-factor of 10 dB, while the NLOS fading can be described by the log-normal distribution with a standard deviation of 0.65 dB. The fading depth ranges from -3 dB to 1.5 dB relative to the mean for LOS, and -0.8 dB to 0.8 dB for NLOS. For the directional antenna pattern, fading in both LOS and NLOS environments follows the Ricean distribution, where the $K$-factor ranges from 7 dB to 17 dB for LOS, and 9 dB to 21 dB for NLOS. and the fading depth varies between -4 dB to 2 dB for both LOS and NLOS environments. 

In terms of the spatial autocorrelation of received signal voltage amplitudes, it follows the sinusoidal-exponential distribution in the LOS environment for the omnidirectional RX antenna and most directional antenna cases, and the exponential distribution for directional RX antenna when the RX antenna moves in a parallel manner with respect to the T-R line. In the NLOS environment, the spatial autocorrelation can be modeled by the exponential distribution for both omnidirectional and directional RX antennas. For the LOS environment, the decorrelation distance ranges from around five wavelengths to 200 wavelengths (2.09 - 81.6 cm), while the NLOS decorrelation distance is between 0.67 and 25 wavelengths (0.27 - 10.2 cm) at 73.5 GHz. It gives rise to the maximum decorrelation distance when the RX antenna points points at a major reflector and moves in a parallel manner with respect to the T-R line, while the minimum decorrelation distance is likely to occur when the RX antenna is pointing roughly to the opposite direction of the TX and without major reflectors. The short correlation distance in most cases is favorable for spatial multiplexing in MIMO, since it allows for uncorrelated spatial data streams to be transmitted from closely-spaced (a fraction to several wavelengths) antennas.

%When the RX antenna moves orthogonally to the T-R line, the autocorrelation is sinusoidally distributed, whereas when the RX antenna moves in a parallel manner with respect to the T-R line and if a dominant path is present, the decreasing rate of the autocorrelation is considerably low. 

%The oscillation distance/period and spatial decay constant are used to represent the autocorrelation, where rapid decorrelation of received voltage amplitudes occurred over a few to ten wavelengths when the RX antenna moves orthogonally to the T-R line. The short correlation distance in general is favorable for spatial multiplexing in MIMO, since it allows for uncorrelated spatial data streams to be transmitted from closely-spaced (a fraction to several wavelengths) antennas.

%if there are two paths interfering with each other, the resultant amplitude first quickly drops to 0 and then becomes anti-correlated as the antenna separation increases from 0 to 30 wavelengths. Further study is encouraged to examine the small-scale statistics for larger antenna separations and temporally over the excess delay for wideband signals. 

%\section*{Appendix}

\ifCLASSOPTIONcaptionsoff
  \newpage
\fi

\bibliographystyle{IEEEtran}
\bibliography{bibliography}

\end{document}